\documentclass{article}

\usepackage{arxiv}

\usepackage[utf8]{inputenc} % allow utf-8 input
\usepackage[T1]{fontenc}    % use 8-bit T1 fonts
\usepackage{hyperref}       % hyperlinks
\usepackage{url}            % simple URL typesetting
\usepackage{booktabs}       % professional-quality tables
\usepackage{amsfonts}       % blackboard math symbols
\usepackage{nicefrac}       % compact symbols for 1/2, etc.
\usepackage{microtype}      % microtypography
\usepackage{lipsum}		% Can be removed after putting your text content
\usepackage{graphicx}
\usepackage{doi}
\usepackage{amsmath}

\title{Optimizing Brain-Computer Interface Performance: Advancing EEG Signals Channel Selection through Regularized CSP and SPEA II Multi-Objective Optimization}

\author{M. Moein Esfahani\\
	Center for Translational Research in Neuroimaging and Data Science (TReNDS)\\
	Georgia State University\\
	Atlanta, Georgia, USA \\
	\texttt{mesfahani1@gsu.edu} \\
	\And
	Hossein Sadati \\
	Department of Mechanical Engineering\\
	K.N Toosi University of Technology\\
	Tehran, Iran \\
	\texttt{sadati@kntu.ac.ir} \\
        \And
	Vince D Calhoun\\
	Tri-Institutional Center for Translational Research in Neuroimaging and Data Science (TReNDS)\\
	Georgia State University, Georgia Institute of Technology, Emory University\\
	Atlanta, Georgia, USA\\
	\texttt{vcalhoun@gsu.edu} \\
}

\date{}

\renewcommand{\shorttitle}{\textit{arXiv} Template}

\hypersetup{
pdftitle={paper Arxiv},
pdfsubject={BCI optimization},
pdfauthor={Moein Esfahani},
pdfkeywords={BCI; EEG; Motor-Imagery; SPEA-II; RCSP; More},
}

\begin{document}
\maketitle

\begin{abstract}
	Brain-computer interface (BCIs) systems and the recording of brain activity has garnered significant attention across a diverse spectrum of applications. Electroencephalography (EEG) signals have emerged as a predominant modality for recording neural electrical activity. Among the methodologies designed for feature extraction from EEG data, the method of regularized common spatial pattern(RCSP) has proven to be an instrumental approach, particularly in the context of Motor Imagery (MI) tasks. RCSP exhibits remarkable efficacy in the discrimination and classification of EEG signals. In the pursuit of optimizing the performance of this method, our research extends to a comparative analysis with conventional CSP techniques, as well as meticulously optimized methodologies designed for similar applications. Notably, we employ the meta-heuristic multi-objective Strength Pareto Evolutionary Algorithm II (SPEA-II) as a pivotal component of our research paradigm. This is a state-of-the-art approach in the selection of an subset of channels from a multi-channel EEG signal with motor imagery tasks. Our main objective is to formulate an optimum channel selection strategy aimed at identifying the most pertinent subset of channels from the multi-dimensional electroencephalogram (EEG) signals. One of the primary ob-jectives inherent to channel selection in the EEG signal analysis pertains to the reduction of the channel count, an approach that enhances user comfort when utilizing gel-based EEG electrode systems. Additionally, within this research, we took benefit of ensemble learning models as a component of our decision-making process. This technique serves to mitigate the challenges asso-ciated with overfitting, especially when confronted with an extensive array of potentially redun-dant EEG channels and data noise. Our findings not only affirm the performance of RCSP in MI-based BCI systems, but also underscore the significance of channel selection strategies and en-semble learning techniques in optimizing the performance of EEG signal classification.
\end{abstract}

% keywords can be removed

\begin{figure}[t]
    \centering
    \includegraphics[width=\textwidth]{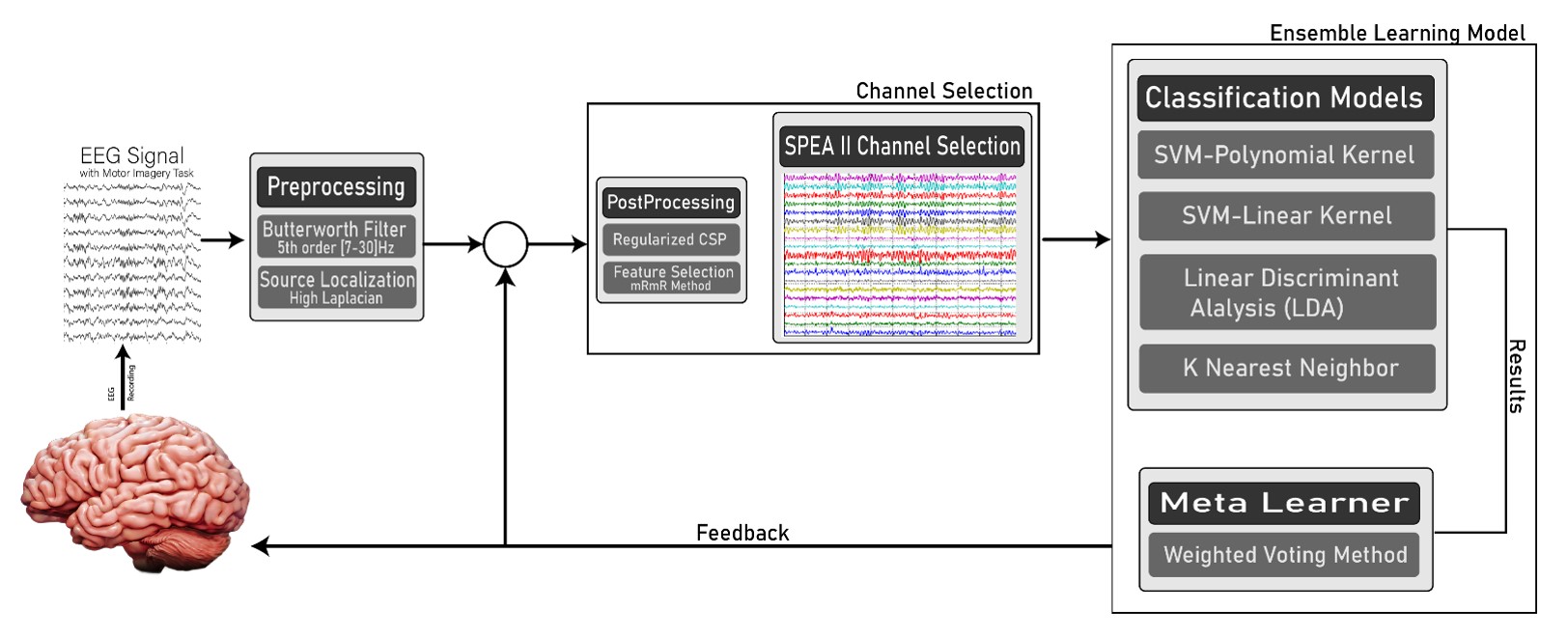}
    \caption{Diagram illustrating the Proposed Channel Selection Method}
    \label{fig:fig1}
\end{figure}

\section{Introduction}

%-----------------------------------------------------------------------------------------------------------------------------------
The evolution of brain-computer interfaces (BCIs) stands as a tailored system that performs at the intersection of neuroscience, computer science, and engineering [1][2], [3]. This combination has given rise to a pioneering technology that enables direct communication between the human brain and external devices, transcending the traditional boundaries of human-machine interaction [4]–[6]. BCIs hold immense potential across a spectrum of applications, ranging from medical rehabilitation to augmenting human capabilities and understanding the brain’s dynamics [2].
The Brain-Computer Interface systems mainly performs in neural recording/decoding and stimulation [3]. Early experimentation involved noninvasive approaches like implementing direct implantation of electrodes into the brain to decipher neural activity with mainly EEG/fNIRS signal recording systems[7], [8] which leverage electroencephalography (EEG) to capture neural signals from the scalp surface[1], [8]–[10].
The advent of BCIs empowering individuals to control external devices through neural [commands mainly with EEG signals…rephrase][1]. This innovation came with profound implications, offering a lifeline for those with motor disabilities and opening avenues for communication restoration. Furthermore, BCIs offered new methods into understanding neural processes, enabling researchers to detect complexities of cognition and mental states [1], [8], [10].
Motor-imagery tasks represent a captivating realm where the human mind's capacity to simulate movement converges with cutting-edge neuro studies. The inception of motor imagery research dates back several years, where pioneers began investigating the neural signatures accompanying imagined actions. This endeavor marked a paradigm shift, transcending traditional notions of motor control and paving the way for novel applications that extend beyond the realm of basic neuroscience [11]–[13].
EEG signals hold the key to unlocking the intricacies of motor imagery. These electrical signals, captured non-invasively from the scalp, offer a window into the brain's activity during cognitive tasks, including motor imagery. By analyzing the spectral patterns and spatiotemporal dynamics of EEG signals, researchers have uncovered distinct neural correlates associated with various motor imagery tasks. These discoveries serve as the foundation for decoding intentions and translating them into tangible actions through BCIs [14]. The realm of optimization, a foundation in various disciplines, continuously seeks solutions to complex problems by navigating a land-scape of possibilities. As problems become increasingly intricate, traditional optimization techniques often grapple with the curse of dimensionality and the need to simultaneously balance conflicting objectives [15], [16]. This predicament gave rise to a transformative innovation – Multi-Objective Metaheuristic Optimizations – an interdisciplinary combination of optimization theory and metaheuristic algorithms [17], [18]. In this study, we use optimization meta-heuristic algorithms and compare the results with our proposed techniques and other algorithms to improve performance [1], [19]–[21].
The journey of Multi-Objective Metaheuristic Optimizations traces its roots to the mid-20th century [15], [22], [23], when researchers recognized that many real-world problems entailed multiple, often competing objectives. Conventional optimization methods struggled to handle the intricate trade-offs inherent in these scenarios, sparking a quest for techniques capable of exploring diverse solution spaces [17], [18]. Metaheuristic algorithms emerged as a class of strategies that sidestepped the rigid constraints of deterministic approaches [16], [24], adopting an iterative, exploratory approach that mimicked natural evolutionary processes like evolution, swarm behavior, and simulated annealing.
The rise of multi-objective optimization was a significant change. Instead of looking for just one best solution, multi-objective optimization aimed to uncover a set of solutions that represented different trade-offs between objectives [1]. This set, known as the Pareto front, offered decision-makers a spectrum of choices, allowing them to make informed decisions that aligned with their preferences and priorities.
In former methodologies, the integration of elitism. was not explicitly incorporated. However, a few years later, the significance of this concept in the context of multi-objective optimization was acknowledged, leading to its incorporation in a new generation of algorithms. This marked the introduction of elitist multi-objective evolu-tionary algorithms such as SPEA by Zitzler[25], [26] and NSGA by Deb [1], [8], [10].These evolutionary strategies underscored the emphasis on preserving elite solutions within the population, thereby enhancing the convergence towards optimal solutions across multiple objectives.
In this study, we present an approach to EEG signal processing and also strategies for mitigating noise and redundancies. We also provided an in-depth exposition of the Regularized CSP technique and its relevance and application within the context of our research. Subsequently, we delve into the SPEA-II and its components, elucidating their roles and significance in our analytical framework. Furthermore, we offer an overview of the dataset employed in our study and the experimental procedures undertaken by our methodologies. Finally, the authors intend to provide readers with a better understanding of the implications, significance, and potential future directions arising from our research efforts.
%-----------------------------------------------------------------------------------------------------------------------------------

\section{Methodology}
\label{sec:headings}

In this study, we introduce our state-of-the-art methodology, as illustrated in Figure 1. Our primary objective is to determine the optimal set of channels using the SPEA-II algorithm. This method operates as a subset of the wrapper approach, focusing on subject-specific channel selection, ensuring that it operates independently for each subject. This tailored approach to channel selection within EEG signal processing for motor imagery tasks offers several advantages, including the reduction of channels to a manageable number, enhancing the comfort of subjects, and reducing setup and electrode fixation times.
An essential consideration in our study is the risk of overfitting in classification out-comes when dealing with an excessive number of redundant EEG recorded channels and signal noise. To address this, we explore five primary strategies for channel selection: filtering, wrapping, embedding, hybrid, and human-based methods[27], [28]. Filtering techniques independently evaluate candidate channel subsets using metrics including distance, dependence, and consistency. These techniques offer advantages such as computational efficiency and independence from specific classification algo-rithms. However, they may exhibit lower accuracy as they don't account for combinations of different channels[1], [7], [8], [10]. 
Wrapper techniques, on the other hand, evaluate the performance of specific classifiers by classifying subsets of features and identifying the subset of channels that per-forms best with a particular classification method. Embedded techniques rely on re-cursive channel elimination. Hybrid approaches combine filtering and wrapper techniques to leverage the strengths of both methodologies[1]. This approach allows us to benefit from independent measures while employing a wrapper method to identify the optimal feature subset. Human-based methods involve domain expertise, where well-trained experts determine which channels are most relevant to specific tasks. For instance, experts in brain physiology may identify channels in the brain's cortex as crucial for motor imagery tasks. In our study, we incorporate this expert knowledge into our algorithm to identify optimal channels within specific brain regions.
Our proposed method predominantly functions as a wrapper technique, employing a specialized classifier to determine the optimal channel subset. This approach aims to strike a balance between accuracy and computational efficiency, contributing to the effectiveness of our channel selection strategy.

\subsection{SPEA-II Algorithm}

SPEA II multi objective optimization algorithm, It is an advanced multi-objective optimization algorithm, first time proposed by Zitzler et al[1]. It introduces a significant advanced method in the domain of Pareto-based optimization. Operating on the foundational principle of Pareto optimization as applied to fitness functions or objectives, SPEA2 not only incorporates selection operations but also like NSGA-II features a state-of-the-art method coupled with an external archiving elite retention mechanism.
In this paper, we delve into a brief introduction of SPEA2, as related to theoretical foundations while attaining more efficient outcomes within the realm of multi-objective algorithms. An enhancement in SPEA2, distinguishing it from its predecessor past version SPEA, is the adoption of an improved fitness assignment. This algorithm with details accounts for the domination relationships of each individual, considering both the number of individuals it dominates and those by which it is dominated. 
Furthermore, SPEA2 integrates a nearest neighbor density estimation technique that intricately guides the search process [25], [26]. This addition empowers the algorithm to navigate the solution space with heightened precision, thereby augmenting the convergence towards Pareto-optimal solutions. Notably, SPEA2 introduces a new ar-chive truncation method that plays a critical role in preserving boundary solutions. This enhancement ensures that solutions at the edge of the Pareto front, pivotal for comprehensive problem understanding, are retained within the solution set. Figure 2 represents a psudocode of SPEA-II.

\begin{figure}
    \centering
    \includegraphics[width=0.4\textwidth]{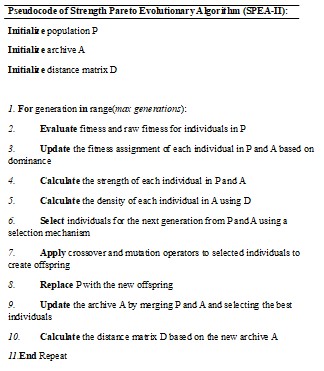} 
    \caption{PseudoCode of SPEA II}
    \label{fig:psudo}
\end{figure}

\subsubsection{Pareto Front Solutions}
In multi-objective optimization, the Pareto front represents a fundamental concept that embodies the trade-offs and compromises between conflicting objectives[17], [18], [25], [26]. It is a collection of solutions where no single solution can be improved in one objective without sacrificing performance in another. With that said, these solutions form the frontier of the solution space, showcasing the optimal compromises that exist among the multiple objectives under consideration. The Pareto front is a crucial tool for decision-makers, as it provides a comprehensive view of the diverse solutions available[1], [7]. By exploring the Pareto front, decision-makers can gain valuable in-sights into the complex interplay of objectives, aiding in the selection of solutions that align most closely with their desired outcomes and objectives.
\begin{table}[h]
    \centering
    \begin{tabular}{|l|l|}
    \hline
    \textbf{Parameters used in SPEA-II algorithm} \\
    \hline
    Iteration & 25 \\
    Population size & 80 \\
    Probability of crossover & 0.75 \\
    Probability of mutation & 0.7 \\
    Type of selection & tournament \\
    \hline
    \end{tabular}
    \caption{SPEA-II Parameters}
    \label{tab:spea_parameters}
\end{table}

\begin{figure}[h]
    \centering
    \includegraphics[width=0.4\textwidth]{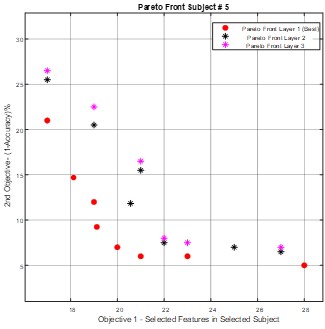} 
    \caption{Visualization of the Pareto Front for Subject Number5}
    \label{fig:fig3}
\end{figure}

\subsubsection{Paragraph}

The origin of the CSP algorithm goes back to the process of extracting discriminative spatial patterns from multichannel EEG signals (mainly with spatial patterns) [2], [3], [29]. The base and foundation of CSP lies in its ability to transform the EEG channels into a new spatial domain [4], [30], [31], where the variance of one class is maximized while that of another class is minimized. the results of the neural activity patterns to certain cognitive states while attenuating unrelated activity.
The benefits of CSP include the reduction of redundant or noisy data. This research in-tends to implement various EEG signal noise reduction techniques other than just CSP, aimed at enhancing signal quality. The CSP was just used as a feature extraction method. Furthermore, this methodology endeavors to prioritize one task over another. However, it is important to note that the algorithm is limited to datasets featuring two tasks. Yet, with some extensions and additional features, we can improve its applica-tion to datasets with more than two tasks [4], [30]–[32].
in Regularized CSP method, the optimization terms need the three parameters $\alpha, \beta and \gamma$. These vectors of parameters, as shown below, are defined based on [33][10], [34] studies:

$\alpha = \{10^{-10}, 10^{-9}, 10^{-8}, 10^{-7}, 10^{-6}, 10^{-5}, 10^{-4}, 10^{-3}, 10^{-2}, 10^{-1}\}$\\
$\beta = \{0.1, 0.2, 0.3, 0.4, 0.5, 0.6, 0.7, 0.8, 0.9\}$\\
$\gamma = \{0.1, 0.2, 0.3, 0.4, 0.5, 0.6, 0.7, 0.8, 0.9\}$

Subsequently, it becomes critical to identify the optimal parameters that can yield peak performance. To achieve this, we endeavor to pinpoint the most effective parameter values. Furthermore, in the context of feature selection from the Regularized Common Spatial Pattern (RCSP) results, the following approach is employed. The process in-volves the utilization of the Minimum Redundancy Maximum Relevance (mRmR) al-gorithm to discern the most pertinent subset of features. It is noteworthy that the RCSP yields N*d trials, where N signifies the number of features extracted through RCSP, and d signifies the number of filters applied to the signal.
With that said, we embark on the selection of the optimal feature subset tailored to each specific subject. The outcomes of this selection process shed light on the precise number of features that comprise the most optimal subset. This strategic approach aids in enhancing the overall performance and effectiveness of our methodology.
CSP uses spatial filters that extremize the following cost function:

\begin{equation}
J(w) = \frac{{w^T C_1 w}}{{w^T C_2 w}} = \frac{{w^T X_1 X_1^T w}}{{w^T X_2 X_2^T w}}
\end{equation}

The signal trial for class I is transformed into a matrix format, where the rows correspond to channels, and the columns correspond to discrete time points. Central to our approach is Eq. (1), which evaluates the cost function, a critical component of our spa-tial covariance matrix for class i. While the initial goal is to maximize this cost function, it is pertinent to acknowledge that conventional Common Spatial Pattern (CSP) methods, despite their widespread use and proven efficacy, come with certain limitations. Among the main drawbacks is a high sensitivity to noise, a factor that can lead to overfitting, particularly when dealing with datasets with a limited number of trials. To overcome these limitations, we adopt the Regularized CSP (denoted as RCSP) as a new method.
The spatial covariance matrix for each of the two classes should be computed. However, under some conditions it may overfit, resulting in low classification accuracy. As a result, the RCSP incorporates an added penalty term in the denominator of the cost function to normalize the findings and reduce the effect of overfitting. The term $\alpha P(w)$  in this cost function represents the penalty term, as follows:

\begin{equation}
J_{P_{1,2}}(w) = \frac{{w^T \tilde{C}_{1,2} w}}{{w^T \tilde{C}_{2,1} w + \alpha P(w)}}
\end{equation}

The task has the estimation of spatial covariance matrices for both of the two distinct classes, which is a base step accomplished through the use of Common Spatial Pattern (CSP) analysis. Nonetheless, when dealing with EEG trial datasets that include noise or a limited number of trials, there arises the potential for overfitting in the computed covariance matrix, which can subsequently lead to suboptimal discrimination results. To address this inherent challenge, we have used an advanced method that applies a penalty term into the cost function, strategically incorporated to normalize the out-comes and mitigate the overfitting effect. This new cost function features the term $\alpha P(w)$ , denoting the penalty term, thoughtfully appended to the denominator of the equation, thus yielding the following formulation:

\begin{equation}
   P(w) = w^T Kw 
\end{equation}
\begin{equation}
    \hat{C}_c = (1-\beta) s_c C_c + \beta G_c
\end{equation}
\begin{equation}
\tilde{C}_c = (1-\gamma) \hat{C}_c + \gamma I 
\end{equation}
\begin{equation}
G_c = \sum_{i \in \Omega} \frac{{N_c^i}}{{N_{t,c}}} C_c^i
\end{equation}
\begin{equation}
s_c = \frac{{N_c}}{{N_{t,c}}} 
\end{equation}

where $\hat{C}_c$ signifies the initial spatial covariance matrix associated with class c, whereas  $\hat{C}_c$ is the regularized estimate, I is the identity matrix, $s_c$ is a constant scaling parameter denoting the identity matrix. Eq. (6) is a generic covariance matrix to estimate the more robust covariance matrix $\hat{C}_c$. The generic matrix obtains its initial values from a given prior data based on subject-to-subject training. The covariance matrix depends on the other recorded subjects. This term applies to the replaced $C1$ and $C2$ used in the CSP transfer function by their regularized estimates.

\begin{equation}
    Z = W^T \cdot x
\end{equation}

\begin{equation}
    f_p = \log \left( \frac{{\text{var}(Z)}}{{\sum \text{var}(Z)}} \right)
\end{equation}

In Eq.(8) we applied the computed weights W on the EEG signals x, giving Z as the mapped EEG trials.\\
Finally, the extracted features are log power features in Eq. (9), where $var(z)$ signifies the variance of $\hat{z}$ over the domain samples. \\ Consequently, from the matrix  $Z=[Z_1,Z_2,Z_3,\dots,Z_n ]^T$, which is the feature matrix, the columns are chosen as follows. The first and last columns are taken, followed by the second and one to the last, followed again by the third and two to the last, and this process is done until all columns are considered. Subsequently, the classification models append on this matrix of features.

\begin{table}[h]
    \centering
    \includegraphics[width=0.5\textwidth]{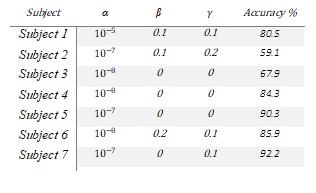} 
    \caption{Regularized CSP Parameters for Subjects}
    \label{fig:tbl2}
\end{table}

\subsection{Source localization with High Laplacian}
In this study, we focused on analyzing EEG signals sourced from a dataset having 64 channels. To enhance the quality of the EEG data, in the present study we employed a source localization High Laplacian spatial filtering method, in contrary to the earlier research by the authors [1], [10]. By taking this approach, we aimed to improve the signal-to-noise ratio (SNR) in order to enhance the performance of EEG signal classification [1], [7], [8], [10]. In other publications, the utilization of common average reference (CAR) and low or large Laplacian spatial filters has been shown to reduce SNR, consequently leading to enhanced EEG signal classification performance [20], [35]. It aimed to facilitate the accentuation of localized neural activity while concurrently mitigating the influence of diffuse signals. This enhancement is achieved by computing the weighted sum of the potentials recorded by neighboring electrodes and then subtracting this sum from the potential measured at the electrode of interest. This process enables the derivation of low Laplacian spatial filters, thereby refining the representation of localized neural activity within the EEG dataset.

\subsection{Digital EEG Signal Filtering}
In our study, we filtered the EEG signals using a 5th order bandpass Butterworth filter in the range of 7-32 Hz.( Unlike prior publication[1], [10], our approach involved a reduction in the filter order and employed a wider filter design) This filtering helped remove unrelated components and noise from the signal, focusing on the frequencies associated with motor imagery, specifically the mu (8-12 Hz) and beta (18-26 Hz) rhythms, which can induce event-related synchronization (ERS) and event-related desynchronization (ERD) phenomena[22], [23], [36]–[38].

\subsection{Ensemble learning Classification}
Ensemble learning models [8], was introduced as a new and powerful approach in machine learning. The ensemble learning consists of the idea of combining multiple individual models, often of diverse types or trained on distinct subsets of data, to make collective predictions that tend to be more accurate and robust than those of any single model. The voting model, as a pivotal component of this ensemble, consolidates the predictions from its constituent models, which could be classifiers like decision trees, support vector machines, or neural networks [37]–[40]. The key idea here is to leverage the wisdom of the crowd: while individual models may have their limitations and biases, their collective decision, reached through a voting mechanism, tends to be more reliable and less susceptible to errors. It often leads to superior predictive performance and enhances the generalizability of the resulting system. In our study, we implemented ensemble learning model. The model for making the final decision and optimizing the performance of our optimization algorithm [16], [24], [41]. This ensemble learning model comprises several submodels as depicted in Figure 1, includes Support Vector Machines (SVM) with both linear and Polynomial Kernels, Linear Discriminant Analysis (LDA), and K-Nearest Neighbor (KNN) with a parameter 'k' set to 6.

\section{Dataset}
\subsection{BCI Competition IV-Dataset1}
This dataset includes EEG signals consists of two distinct classes of motor imagery, meticulously recorded via the BrainAmp MR amplifier system (Brain Products GmbH, Munich, Germany)[42]–[45], combined with the utilization of Ag/AgCl electrode caps provided by EASYCAP GmbH. To discuss further, this dataset meticulously incorporates a total of 100 trials exclusively dedicated to motor imagery linked with right-hand movements, supplemented by an equivalent count of 100 trials directed towards the contemplation of left-hand motor imagery. The data collection procedure was performed with the active participation of four healthy human subjects, appended by the involvement of three computationally generated virtual subjects. The data acquisition process involved the recording of neural signals from precisely 59 EEG locations, positioned across the scalp to densely cover sensorimotor regions of interest. The collected signals underwent a comprehensive bandpass filtering operation, expertly constraining them within the frequency bandwidth of 0.05 to 200 Hz. Following this, the neural data was digitized at an impressive sampling rate of 1000 Hz, with precision level of 16 bits (0.1 uV). In this study, we Notably take advantage of the sophisticated Chebyshev Type II filter to judiciously down-sample the dataset to a more operationally efficient sampling rate of 100 Hz[1], [8], [10], [19].

\begin{figure}[h]
    \centering
    \includegraphics[width=0.5\textwidth]{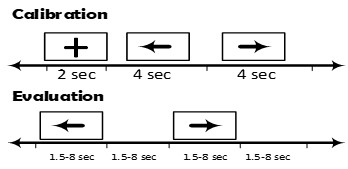} 
    \caption{Visualization of a Single EEG Trial in described Dataset}
    \label{fig:fig4}
\end{figure}

\subsection{Signal Segmentation}
Each participating subject engaged with three distinct motor imagery tasks. These tasks intricately associated with motor imagery involving the left hand, the right hand, or the right foot. To facilitate comprehensive analysis and interpretation, the mentioned dataset was thoughtfully segregated into two principal segments, denoted as the 'calibration' and 'evaluation' phases. Better visulalization in Figure.4. Calibration Data: Within the initial phase of the experiment, specifically during the initial two runs, participants were exposed to four-second visual cues that assumed the form of directional arrows, judiciously pointing either left, right, or down, meticulously presented on a screen. Evaluation Data: Subsequent to the calibration phase, four additional experimental runs were conscientiously executed, with the data from four of these runs being thoughtfully incorporated into the dataset under scrutiny. To cue the subjects into motor imagery tasks, we employed soft acoustic stimuli, manifesting as spoken words such as 'left,' 'right,' and 'foot.' The duration of each motor imagery task was thoughtfully designed to span a range of temporal periods, encompassing intervals that varied from a succinct 1.5 seconds to a more protracted 8 seconds. The demarcation of the termination of each motor imagery phase was aptly conveyed through the cue 'stop,' further embellishing the temporal diversity inherent to the dataset.

\section{Results}
The PSEA-II algorithm was employed in this study to select the optimal channel subset. A Pareto front, depicted in Figure 3 for subject 5, showcases the set of optimal solutions generated by the multi-objective evolutionary algorithm. The Pareto front serves as a valuable tool for decision-making by balancing the trade-off between the number of selected channels and accuracy. Although presented in a two-dimensional format in our study, the Pareto front can manifest in higher dimensions for more complex opti-mization problems. In figure 3, multiple layers of results has depicted. Also, to illus-trate as a convex figure, we employed the data with 1-accuracy. As the result, as the curve of the same legend goes down, the higher accuracy with achieve.\\
The PSEA-II algorithm was executed for 25 iterations, varying by subject, with a population size of 80\%, a crossover percentage of 75, a mutation percentage of 70\%, and a total of 1374 function evaluations. These parameters are also detailed in Table 3. The resulting accuracies, along with the number of selected channels for each subject, are summarized in Table 2. Figure 5 provides a visual representation of electrode locations on the subjects' scalps, offering multiple channel selection options.\\
Notably, in every row, the first circle in Figure 5 represents our secondary objective: minimizing the number of channels at the expense of accuracy, while the last circle signifies our optimal choice, aiming for the highest accuracy with the fewest channels. However, it's observed that the second and third layers of the Pareto front exhibit lower accuracies compared to the initial layer.

\begin{figure}[h]
    \centering
    \includegraphics[width=\textwidth]{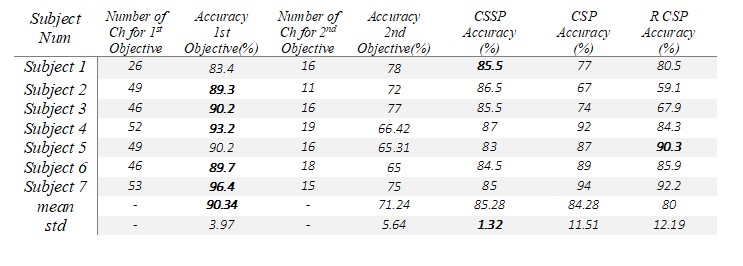} 
    \caption{Discussion of the Number of Optimal Channels Selected and Accuracy Across Various Proposed Algorithms}
    \label{fig:tbl2}
\end{figure}

\section{Discussion}
To validate the performance of our proposed Regularized CSP algorithm, a comparative study against recently published results in the introduction was conducted. Bold entries in Table 2 denote the optimal and highest results. Remarkably, our algorithm demonstrated superior accuracy performance across most subjects, with exceptions for subjects 1 and 5. These results were validated using 10-fold cross-validation, ensuring robustness and reliability.\\
Furthermore, Figure 5 presents the optimal channel selections for all seven subjects, aligning with our primary objective of channel minimization while striving to maximize accuracy. Table 2 conveniently summarizes the accuracy levels and corresponding channel counts selected, providing a comprehensive overview of the algorithm's performance across different subjects.\\
Overall, the utilization of the PSEA-II algorithm in conjunction with the Regularized CSP algorithm showcases promising results in optimizing channel selection for EEG data classification. The ability to balance accuracy with channel reduction offers practical implications in real-world applications where resource constraints or processing efficiency are crucial factors. Further exploration and validation of these methodologies across diverse datasets and experimental setups could provide deeper insights and potentially enhance their applicability in various EEG-based applications.

\section{Conclusion}
This study primarily centers on the utilization of the Strength Pareto Evolutionary Algorithm (SPEA-II) to derive the most optimal channel subset, a critical aspect in the development of a high-performance Brain-Computer Interface (BCI) system. Designing an effective BCI system entails a formidable challenge, primarily stemming from the necessity to identify the optimal feature set. This endeavor is optimized by the computational time complexity of the proposed methods and the pursuit of attaining the highest classification accuracy. In this study for efficient classification and feature extraction, we have introduced the Regularized Common Spatial Pattern (RCSP) method and conducted a comparative analysis against alternative methodologies. In the context of this study, our aim was to ascertain the ideal channel configuration for the BCI Competition IV dataset 1. To achieve this, we used an ensemble learning classification model including multiple kernels to perform EEG trial classification. Our findings underscore the efficacy of Regularized CSP, a cutting-edge CSP variant enriched with a penalty term, which consistently delivered superior accuracy outcomes. In future studies, the inclusion of diverse classification techniques offers potential for further exploration, affording opportunities for result comparison. Additionally, there exists the prospect of assessing the outcomes of alternative evolutionary applications, such as MOPSO or MOEAD, as a new method of multi-objective evolutionary approach.

\begin{figure}[p] 
    \centering
    \includegraphics[width=\textwidth,height=\textheight,keepaspectratio]{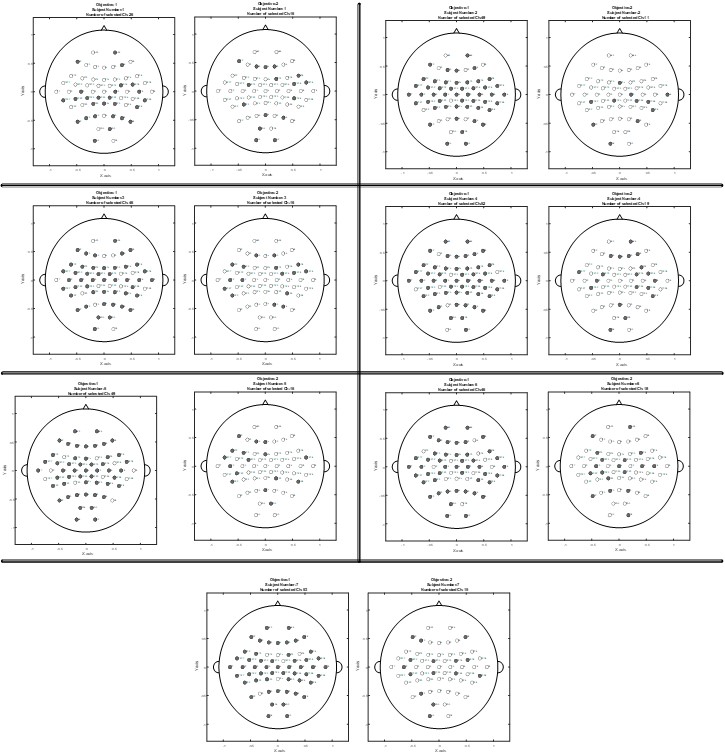} 
    \caption{Optimal Selected Channels per specific Subjects}
    \label{fig:fig5}
\end{figure}

\clearpage
\bibliographystyle{unsrtnat}
%\bibliography{references}  %%% Uncomment this line and comment out the ``thebibliography'' section below to use the external .bib file (using bibtex) .

%%% Uncomment this section and comment out the \bibliography{references} line above to use inline references.

\end{document}